\documentclass[aps,prb,reprint,showpacs,superscriptaddress]{revtex4-1}
\usepackage[dvipdfmx]{graphicx}

\usepackage{amsmath,amssymb}
\usepackage{color}

\usepackage{bm}

\def\v#1{\bm{#1}}			
\def\vr{\v{r}} 					
\def\vp{\v{p}} 					
\def\vq{\v{q}} 					
\def\vk{\v{k}} 					

\def\la{\langle}
\def\ra{\rangle}

\usepackage{comment}
\usepackage{color}
\usepackage[normalem]{ulem}
\usepackage{cancel}

\begin{document}

\title{Ferromagnetic resonance modulation in $d$-wave superconductor/ferromagnetic insulator bilayer systems}

\author{Yuya Ominato}
\affiliation{%
Kavli Institute for Theoretical Sciences, University of Chinese Academy of Sciences, Beijing 100190, China.
}%
\author{Ai Yamakage}
\affiliation{Department of Physics, Nagoya University, Nagoya 464-8602, Japan}
\author{Takeo Kato}
\affiliation{Institute for Solid State Physics, The University of Tokyo, Kashiwa 277-8581, Japan}
\author{Mamoru Matsuo }
\affiliation{%
Kavli Institute for Theoretical Sciences, University of Chinese Academy of Sciences, Beijing 100190, China.
}%
\affiliation{%
CAS Center for Excellence in Topological Quantum Computation, University of Chinese Academy of Sciences, Beijing 100190, China
}%
\affiliation{%
Advanced Science Research Center, Japan Atomic Energy Agency, Tokai 319-1195, Japan
}%
\affiliation{%
RIKEN Center for Emergent Matter Science (CEMS), Wako, Saitama 351-0198, Japan
}%

\date{\today}

\begin{abstract}
We investigate ferromagnetic resonance (FMR) modulation in $d$-wave superconductor (SC)/ferromagnetic insulator (FI) bilayer systems theoretically.
The modulation of the Gilbert damping in these systems reflects the existence of nodes in the $d$-wave SC and shows power-law decay characteristics within the low-temperature and low-frequency limit.
Our results indicate the effectiveness of use of spin pumping as a probe technique to  determine the symmetry of unconventional SCs with high sensitivity for nanoscale thin films.
\end{abstract}

\maketitle 
\section{Introduction}

Spin pumping (SP)~\cite{tserkovnyak2002enhanced,hellman2017interface} is a versatile method that can be used to generate spin currents at magnetic junctions.
While SP has been used for spin accumulation in various materials in the field of spintronics~\cite{Zutic2004,Tsymbal2019}, it has recently been recognized that SP can also be used to detect spin excitation in nanostructured materials~\cite{han2020spin}, including magnetic thin films~\cite{Qiu2016}, two-dimensional electron systems~\cite{Ominato2020a,Ominato2020b,yama2021spin}, and magnetic impurities on metal surfaces~\cite{yamamoto2021}. Notably, spin excitation detection using SP is sensitive even for such nanoscale thin films for which detection by conventional bulk measurement techniques such as nuclear magnetic resonance and neutron scattering experiment is difficult.

Recently, spin injection into $s$-wave superconductors (SCs) has been a subject of intensive study both theoretically~\cite{inoueSpinPumpingSuperconductors2017,taira2018spin,Kato2019,silaevFinitefrequencySpinSusceptibility2020,silaevLargeEnhancementSpin2020,vargas2020injection,Vargas2020,Ojajarvi2020,Simensen2021,Holm2021} and experimentally~\cite{bellSpinDynamicsSuperconductorFerromagnet2008,wakamura2015quasiparticle,jeonEnhancedSpinPumping2018,yaoProbeSpinDynamics2018,liPossibleEvidenceSpinTransfer2018,Umeda2018,jeonEffectMeissnerScreening2019,jeonAbrikosovVortexNucleation2019,jeon2019exchange,Rogdakis2019,golovchanskiyMagnetizationDynamicsProximityCoupled2020,zhaoExploringContributionTrapped2020,Muller2021,Yao2021}.
While the research into spin transport in $s$-wave SC/magnet junctions is expected to see rapid development, expansion of the development targets toward unconventional SCs represents a fascinating research direction.
Nevertheless, SP into unconventional SCs has only been considered in a few recent works~\cite{ominato2021detection,johnsen2021magnon}.
In particular, SP into a $d$-wave SC, which is one of the simplest unconventional SCs that can be realized in cuprate SCs~\cite{RevModPhys.72.969}, has not been studied theoretically to the best of our knowledge, although experimental SP in a $d$-wave SC has been reported recently~\cite{carreira2020spin}.

\begin{figure}[t]
\begin{center}
\includegraphics[width=1\hsize]{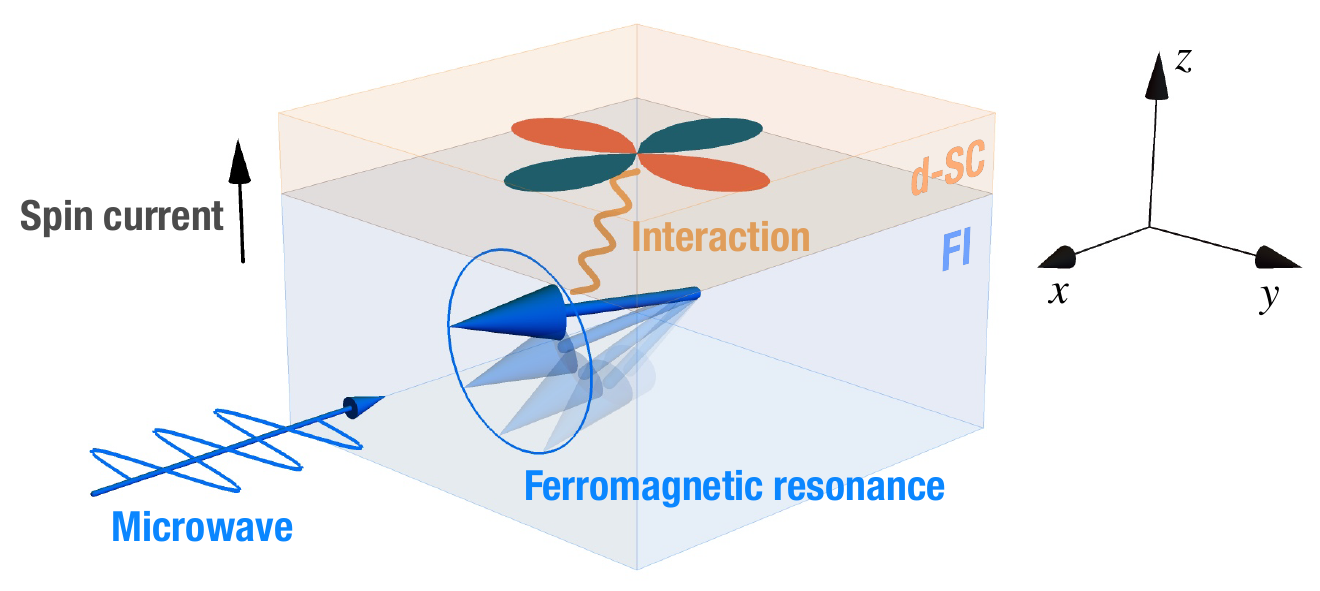}
\end{center}
\caption{
Schematic of the $d$-wave SC/FI bilayer system. The two-dimensional $d$-wave SC is placed on the FI. Precessional motion of the magnetization is induced by microwave irradiation. The spins are injected and the magnetization dynamics are modulated because of the interface magnetic interaction.
}
\label{fig_system}
\end{figure}

In this work, we investigate SP theoretically in a bilayer magnetic junction composed of a $d$-wave SC and a ferromagnetic insulator (FI), as shown in Fig.~\ref{fig_system}. 
We apply a static magnetic field along the $x$ direction and consider the ferromagnetic resonance (FMR) experiment of the FI induced by microwave irradiation.
In this setup, the FMR linewidth is determined by the sum of the intrinsic contribution made by the Gilbert damping of the bulk FI and the interface contribution, which originates from the spin transfer caused by exchange coupling between the $d$-wave SC and the FI.
We then calculate the interface contribution to the FMR linewidth, which is called the modulation of the Gilbert damping hereafter, using microscopic theory based on the second-order perturbation~\cite{Ohnuma2014,ohnuma2017theory,Matsuo2018}.
We show that the temperature dependence of the modulation of the Gilbert damping exhibits a coherent peak below the transition temperature that is weaker than that of $s$-wave SCs~\cite{inoueSpinPumpingSuperconductors2017,Kato2019,silaevFinitefrequencySpinSusceptibility2020,silaevLargeEnhancementSpin2020}.
We also show that because of the existence of nodes in the $d$-wave SCs, the FMR linewidth enhancement due to SP remains even at zero temperature.

The paper is organized as follows. In Sec.~\ref{sec_model}, we introduce the model Hamiltonian of the SC/FI bilayer system. In Sec.~\ref{sec_formulation}, we present the formalism to calculate the modulation of the Gilbert damping. In Sec.~\ref{sec_modulation}, we present the numerical results and explain the detailed behavior of the modulation of the Gilbert damping. In Sec.~\ref{sec_discussion}, we briefly discuss the relation to other SC symmetries, the proximity effect, and the difference between $d$-wave SC/FI junctions and $d$-wave SC/ferromagnetic metal junctions.
We also discuss the effect of an effective Zeeman field due to the exchange coupling.
In Sec.~\ref{sec_conclusion}, we present our conclusion and future perspectives.

\section{Model}
\label{sec_model}

The model Hamiltonian of the SC/FI bilayer system $H$ is given by
\begin{align}
    H=H_{\mathrm{FI}}+H_{d\mathrm{SC}}+H_{\mathrm{T}}.
\end{align}
The first term $H_{\mathrm{FI}}$ is the ferromagnetic Heisenberg model, which is given by
\begin{align}
    H_{\mathrm{FI}} =
    &-\mathcal{J}\sum_{\la i,j\ra}\bm{S}_i\cdot\bm{S}_j
    -\hbar\gamma h_{\mathrm{dc}}\sum_{j}S^x_j, 
\end{align}
where $\mathcal{J}>0$ is the exchange coupling constant, $\la i,j\ra$ represents summation over all the nearest-neighbor sites, $\bm{S}_j$ is the localized spin at site $j$ in the FI, $\gamma$ is the gyromagnetic ratio, and $h_{\mathrm{dc}}$ is the static magnetic field.
The localized spin $\boldsymbol{S}_j$ is described as shown using the bosonic operators $b_j$ and $b_j^\dagger$ of the Holstein-Primakoff transformation\cite{holstein1940field}
\begin{align}
    &S^+_j=S^y_j+iS^z_j=\left(2S-b_j^\dagger b_j\right)^{1/2}b_j, \\
    &S^-_j=S^y_j-iS^z_j=b_j^\dagger\left(2S-b_j^\dagger b_j\right)^{1/2}, \\
    &S^x_j=S-b_j^\dagger b_j,
\end{align}
where we require $[b_i,b_j^\dagger]=\delta_{i,j}$ to ensure that $S^+_j$, $S^-_j$, and $S_j^x$ satisfy the commutation relation of angular momentum.
The deviation of $S^x_j$ from its maximum value $S$ is quantified using the boson particle number.
It is convenient to represent the bosonic operators in the reciprocal space as follows
\begin{align}
    b_{\vk}=\frac{1}{\sqrt{N}}\sum_je^{-i\vk\cdot\vr_j}b_j, \hspace{3mm}
    b_{\vk}^\dagger=\frac{1}{\sqrt{N}}\sum_je^{i\vk\cdot\vr_j}b_j^\dagger,
\end{align}
where $N$ is the number of sites.
The magnon operators with wave vector $\vk=(k_x,k_y,k_z)$ satisfy $[b_{\vk},b_{\vk^\prime}^\dagger]=\delta_{\vk,\vk^\prime}$.
Assuming that the deviation is small, i.e., that $\la b_j^\dagger b_j\ra/S\ll1$, the ladder operators $S_j^\pm$ can be approximated as $S_j^+\approx(2S)^{1/2}b_j$ and $S_j^-\approx(2S)^{1/2}b_j^\dagger$, which is called the spin-wave approximation.
The Hamiltonian $H_{\mathrm{FI}}$ is then written as
\begin{align}
    H_{\mathrm{FI}}\approx
    &\sum_{\vk}\hbar\omega_{\vk}b_{\vk}^\dagger b_{\vk},
\end{align}
where we assume a parabolic dispersion  $\hbar\omega_{\vk}=\mathcal{D}k^2+\hbar\gamma h_{\mathrm{dc}}$ with a spin stiffness constant $\mathcal{D}$ and the constant terms are omitted.

The second term $H_{d\mathrm{SC}}$ is the mean-field Hamiltonian for the two-dimensional $d$-wave SC, and is given by
\begin{align}
    H_{d\mathrm{SC}}=\sum_{\vk}(c_{\vk\uparrow}^\dagger,c_{-\vk\downarrow})
        \begin{pmatrix}
            \xi_{\vk}    && \Delta_{\vk} \\
            \Delta_{\vk} && -\xi_{\vk}
        \end{pmatrix}
        \begin{pmatrix}
            c_{\vk\uparrow} \\
            c_{-\vk\downarrow}^\dagger
        \end{pmatrix},
        \label{eq_Hsc}
\end{align}
where $c_{\vk\sigma}^\dagger$ and $c_{\vk\sigma}$ denote the creation and annihilation operators, respectively, of the electrons with the wave vector $\vk=(k_x,k_y)$ and the $x$ component of the spin $\sigma=\uparrow,\downarrow$, and $\xi_{\vk}=\hbar^2k^2/2m-\mu$ is the energy of conduction electrons measured from the chemical potential $\mu$.
We assume that the $d$-wave pair potential has the form $\Delta_{\vk} = \Delta \cos2\phi_{\vk}$ with the phenomenological temperature dependence
\begin{align}
	\Delta=1.76k_{\mathrm{B}}T_{\mathrm{c}}\tanh\left(1.74\sqrt{\frac{T_{\mathrm{c}}}{T}-1}\right),
\label{Delta_T}
\end{align}
where $\phi_{\vk}=\arctan(k_y/k_x)$ denotes the azimuth angle of $\boldsymbol{k}$.
Using the Bogoliubov transformation given by
\begin{align}
    &\begin{pmatrix}
        c_{\vk\uparrow} \\
        c_{-\vk\downarrow}^\dagger
    \end{pmatrix}
    =
    \begin{pmatrix}
        u_{\vk} && -v_{\vk} \\
        v_{\vk} && u_{\vk}
    \end{pmatrix}
    \begin{pmatrix}
        \gamma_{\vk\uparrow} \\
        \gamma_{-\vk\downarrow}^\dagger
    \end{pmatrix},
    \label{eq_bogoliubov_transformation}
\end{align}
where $\gamma_{\vk\sigma}^\dagger$ and $\gamma_{\vk\sigma}$ denote the creation and annihilation operators of the Bogoliubov quasiparticles, respectively, and $u_{\vk}$ and $v_{\vk}$ are given by
\begin{align}
    u_{\vk}=\sqrt{\frac{E_{\vk}+\xi_{\vk}}{2E_{\vk}}}, \hspace{2mm}
    v_{\vk}=\sqrt{\frac{E_{\vk}-\xi_{\vk}}{2E_{\vk}}},
\end{align}
with the quasiparticle energy $E_{\vk}=\sqrt{\xi_{\vk}^2+\Delta_{\vk}^2}$, the mean-field Hamiltonian can be diagonalized as
\begin{align}
    H_{d\mathrm{SC}}
    =
    \sum_{\vk}(\gamma_{\vk\uparrow}^\dagger,\gamma_{-\vk\downarrow})
        \begin{pmatrix}
            E_{\vk} && 0 \\
            0 && -E_{\vk}
        \end{pmatrix}
        \begin{pmatrix}
            \gamma_{\vk\uparrow} \\
            \gamma_{-\vk\downarrow}^\dagger
        \end{pmatrix}.
\end{align}
The density of states of the $d$-wave SC is given by \cite{coleman2015}
\begin{align}
    D(E)/D_{\mathrm{n}}
    =
    \mathrm{Re}
    \left[
        \frac{2}{\pi}K\left(\frac{\Delta^2}{E^2}\right)
    \right],
\end{align}
where $D_{\mathrm{n}}=Am/2\pi\hbar^2$ is the density of states per spin of the normal state, $A$ is the system area, and $K(x)$ is the complete elliptic integral of the first kind in terms of the parameter $x$, where
\begin{align}
    K(x)=\int^{\pi/2}_0\frac{d\phi}{\sqrt{1-x\cos^2\phi}}.
\end{align}
$D(E)$ diverges at $E/\Delta=1$ and decreases linearly when $E/\Delta\ll1$ because of the nodal structure of $\Delta_{\vk}$.
The density of states for an $s$-wave SC, in contrast, has a gap for $|E| < \Delta$. This difference leads to distinct FMR modulation behaviors, as shown below.

\begin{figure}
\begin{center}
\includegraphics[width=1\hsize]{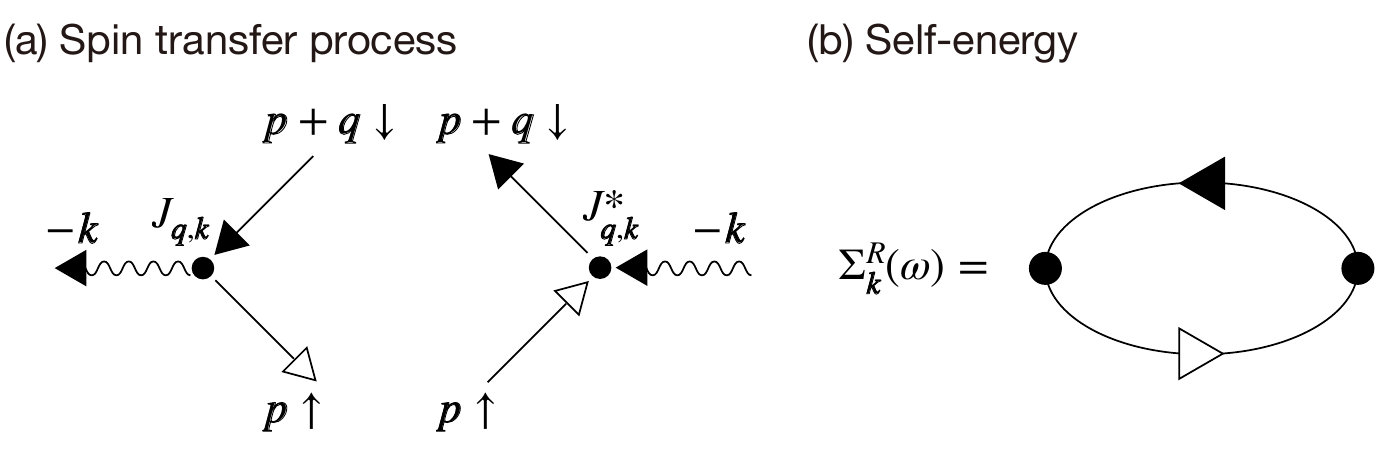}
\end{center}
\caption{
(a) Diagrams of the bare vertices of the spin transfer processes at the interface. (b) Self-energy within the second-order perturbation.}
\label{fig_diagram}
\end{figure}

The third term $H_{\mathrm{T}}$ describes the spin transfer between the SC and the FI at the interface
\begin{align}
    H_{\mathrm{T}}
    =
    \sum_{\vq,\vk}
    \left(
        J_{\vq,\vk}\sigma^+_{\vq}S^-_{\vk} +
        J_{\vq,\vk}^\ast\sigma^-_{-\vq}S^+_{-\vk}
    \right),
\end{align}
where $J_{\vq,\vk}$ is the matrix element of the spin transfer processes, and $\sigma^{\pm}_{\vq}=(\sigma^y_{\vq}\pm i\sigma^z_{\vq})/2$ and  $S^{\pm}_{\vk}=S^y_{\vk}\pm iS^z_{\vk}$ are the Fourier components of the ladder operators and are given by
\begin{align}
    &\sigma^+_{\vq}
    =
    \sum_{\vp}c_{\vp\uparrow}^\dagger c_{\vp+\vq\downarrow}, \hspace{2mm}
    \sigma^-_{-\vq}
    =
    \sum_{\vp}c_{\vp+\vq\downarrow}^\dagger c_{\vp\uparrow}, \\
    &S^-_{-\vk}\approx(2S)^{1/2}b_{\vk}^\dagger, \hspace{2mm}
    S^+_{\vk}\approx(2S)^{1/2}b_{\vk}.
\end{align}
Using the expressions above, $H_{\mathrm{T}}$ can be written as
\begin{align}
    H_{\mathrm{T}}
    \approx
    \sqrt{2S}
    \sum_{\vp,\vq,\vk}
    \left(
        J_{\vq,\vk}
        c_{\vp\uparrow}^\dagger c_{\vp+\vq\downarrow}
        b_{-\vk}^\dagger
        +
        J_{\vq,\vk}^\ast
        c_{\vp+\vq\downarrow}^\dagger c_{\vp\uparrow}
        b_{-\vk}
    \right).
\end{align}
The first (second) term describes a magnon emission (absorption) process accompanying an electron spin-flip from down to up (from up to down).
A diagrammatic representation of the interface interactions is shown in Fig.~\ref{fig_diagram}~(a).

In this work, we drop a diagonal exchange coupling at the interface, whose Hamiltonian is given as
\begin{align}
H_{\mathrm{Z}}=\sum_{{\bm q},{\bm k}} 
J_{\vq,\vk} \sigma^x_{\vq}S^x_{\vk} .
\label{eq:Hzzcoupling}
\end{align}
This term does not change the number of magnons in the FI and induces an effective Zeeman field on electrons in the two-dimensional $d$-wave SC.
We expect that this term does not affect our main result because the coupling strength is expected to be much smaller than the superconducting gap and the microwave photon energy.
We will discuss this effect in Sec.~\ref{sec_discussion} briefly.

\section{Formulation}
\label{sec_formulation}

The coupling between the localized spin and the microwave is given by
\begin{align}
    V(t)
    &=-\hbar\gamma h_{\mathrm{ac}}\sum_i\left(S^y_i\cos\omega t-S^z_i\sin\omega t\right),
\end{align}
where $h_{\mathrm{ac}}$ is the amplitude of the transverse oscillating magnetic field with frequency $\omega$.
The microwave irradiation induces the precessional motion of the localized spin.
The Gilbert damping constant can be read from the retarded magnon propagator defined by
\begin{align}
    G^R_{\vk}(t)
    =
    \frac{1}{i\hbar}\theta(t) \la [S^+_{\vk}(t), S^-_{-\vk}(0)] \ra ,
\end{align}
where $\theta(t)$ is a step function.
Second-order perturbation calculation of the magnon propagator with respect to the interface interaction was performed and the expression of the self-energy was derived in the study of SP \cite{Ohnuma2014,ohnuma2017theory,Matsuo2018}.
Following calculation of the second-order perturbation with respect to $J_{\vq,\vk}$, the Fourier transform of the retarded magnon propagator is given by
\begin{align}
    G^R_{\vk}(\omega)
    =
    \frac{2S/\hbar}{\omega-\omega_{\vk}+i\alpha\omega-(2S/\hbar)\Sigma^R_{\vk}(\omega)},
\end{align}
where $\alpha$ is the intrinsic Gilbert damping constant that was introduced phenomenologically \cite{kasuyaRelaxationMechanismsFerromagnetic1961,cherepanovSagaYIGSpectra1993,jinTemperatureDependenceSpinwave2019}.
The diagram of the self-energy $\Sigma_{\boldsymbol{k}}^R(\omega)$ is shown in Fig.~\ref{fig_diagram}~(b).
From the expressions given above, the modulation of the Gilbert damping constant is given by
\begin{align}
    \delta\alpha
    =
    -\frac{2S\,\mathrm{Im}\,\Sigma^R_{\vk=\bm{0}}(\omega)}{\hbar\omega}.
\end{align}

Within the second-order perturbation, the self-energy is given by
\begin{align}
    \Sigma_{\vk}^R(\omega)
    =
    -\sum_{\vq}|J_{\vq,\vk}|^2\chi^R_{\vq}(\omega),
    \label{Sigma}
\end{align}
where $\chi^R_{\vq}(\omega)$ represents the dynamic spin susceptibility of the $d$-wave SC defined by
\begin{align}
    &\chi^{R}_{\vq}(\omega)
    =
    -\frac{1}{i\hbar}\int dte^{i(\omega+i0)t}
    \theta(t)\la[\sigma^+_{\vq}(t),\sigma^-_{-\vq}(0)]\ra.
    \label{eq_chi_qw_def}
\end{align}
Substituting the ladder operators in terms of the Bogoliubov quasiparticle operators into the above expression and performing a straightforward calculation, we then obtain \cite{coleman2015}
\begin{widetext}
\begin{align}
    \chi^{R}_{\vq}(\omega)=
        -\sum_{\vp}
        \sum_{\lambda=\pm1}
        \sum_{\lambda^\prime=\pm1}
        &\left(
            \frac
            {(\xi_{\vp}+\lambda E_{\vp})(\xi_{\vp+\vq}+\lambda^\prime E_{\vp+\vq})+\Delta_{\vp}\Delta_{\vp+\vq}}
            {4\lambda E_{\vp}\lambda^\prime E_{\vp+\vq}}
        \right)\frac
        {f(\lambda E_{\vp})-f(\lambda^\prime E_{\vp+\vq})}
        {\lambda E_{\vp}-\lambda^\prime E_{\vp+\vq}+\hbar\omega+i0},
        \label{eq_chi_qw}
\end{align}
\end{widetext}
where $f(E)=1/(e^{E/k_{\mathrm{B}}T}+1)$ is the Fermi distribution function.

In this paper, we focus on a rough interface modeled in terms of the mean $J_1$ and variance ${J_2}^2$ of the distribution of $J_{\vq,\vk}$ (see Appendix~\ref{roughness} for detail). 
	The configurationally averaged coupling constant is given by
\begin{align}
 |J_{\boldsymbol{q}, \boldsymbol{k}=\boldsymbol{0}}|^2 = {J_1}^2 \delta_{\vq,\bm{0}} + {J_2}^2.
 \label{Jqk}
\end{align}
In this case, $\delta\alpha$ is written as
\begin{align}
    \delta\alpha
    =
    \frac{2S{J_1}^2}{\hbar\omega}
    \mathrm{Im} \,
    \chi^R_{\vq=\bm{0}}(\omega)
    +
    \frac{2S{J_2}^2}{\hbar\omega}
    \sum_{\boldsymbol{q}}
    \mathrm{Im} \,
    \chi^R_{\vq}(\omega).
\label{delta_alpha}
\end{align}
The first term represents the momentum-conserved spin-transfer processes, which vanish as directly verified from Eq.~(\ref{eq_chi_qw}).
This vanishment always occurs in spin-singlet SCs, including $s$ and $d$-wave SCs, since the spin is conserved \cite{coleman2015}.
Consequently, the enhanced Gilbert damping is contributed from spin-transfer processes induced by the roughness proportional to the variance ${J_2}^2$
\begin{align}
	\delta\alpha
	=
	\frac{2S{J_2}^2}{\hbar\omega}
	\sum_{\vq}
	\mathrm{Im}\,
	\chi^R_{\vq}(\omega).
	\label{eq_delta_alpha_chi_loc}
\end{align}
The wave number summation can be replaced as
\begin{align}
    \sum_{\vq}(\cdots)
    \to
    \frac{D_{\mathrm{n}}}{2\pi}\int^{\infty}_{-\infty}d\xi\int^{2\pi}_0d\phi(\cdots).
\end{align}
Changing the integral variable from $\xi$ to $E$ and substituting Eq.\ (\ref{eq_chi_qw}) into Eq.\ (\ref{eq_delta_alpha_chi_loc}), we finally obtain
\begin{align}
    \delta\alpha
    =
    &\frac{2\pi S {J_2}^2 D_{\mathrm{n}}^2}{\hbar\omega}
    \int^\infty_{-\infty}dE
    [f(E)-f(E+\hbar\omega)] \notag \\
    &\times
    {\mathrm{Re}}
    \left[
        \frac{2}{\pi}K\left(\frac{\Delta^2}{E^2}\right)
    \right]
    {\mathrm{Re}}
    \left[
        \frac{2}{\pi}K\left(\frac{\Delta^2}{(E+\hbar\omega)^2}\right)
    \right].
\end{align}
Note that the coherence factor vanishes in the above expression by performing the angular integral.
The enhanced Gilbert damping in the normal state is given by
\begin{align}
    \delta\alpha_{\mathrm{n}}=2\pi S {J_2}^2D_{\mathrm{n}}^2,
\end{align}
for the lowest order of $\omega$. This expression means that $\delta\alpha$ is proportional to the product of the spin-up and spin-down densities of states at the Fermi level \cite{Ominato2020a}.

\section{Gilbert damping modulation}
\label{sec_modulation}

\begin{figure}
\begin{center}
\includegraphics[width=1\hsize]{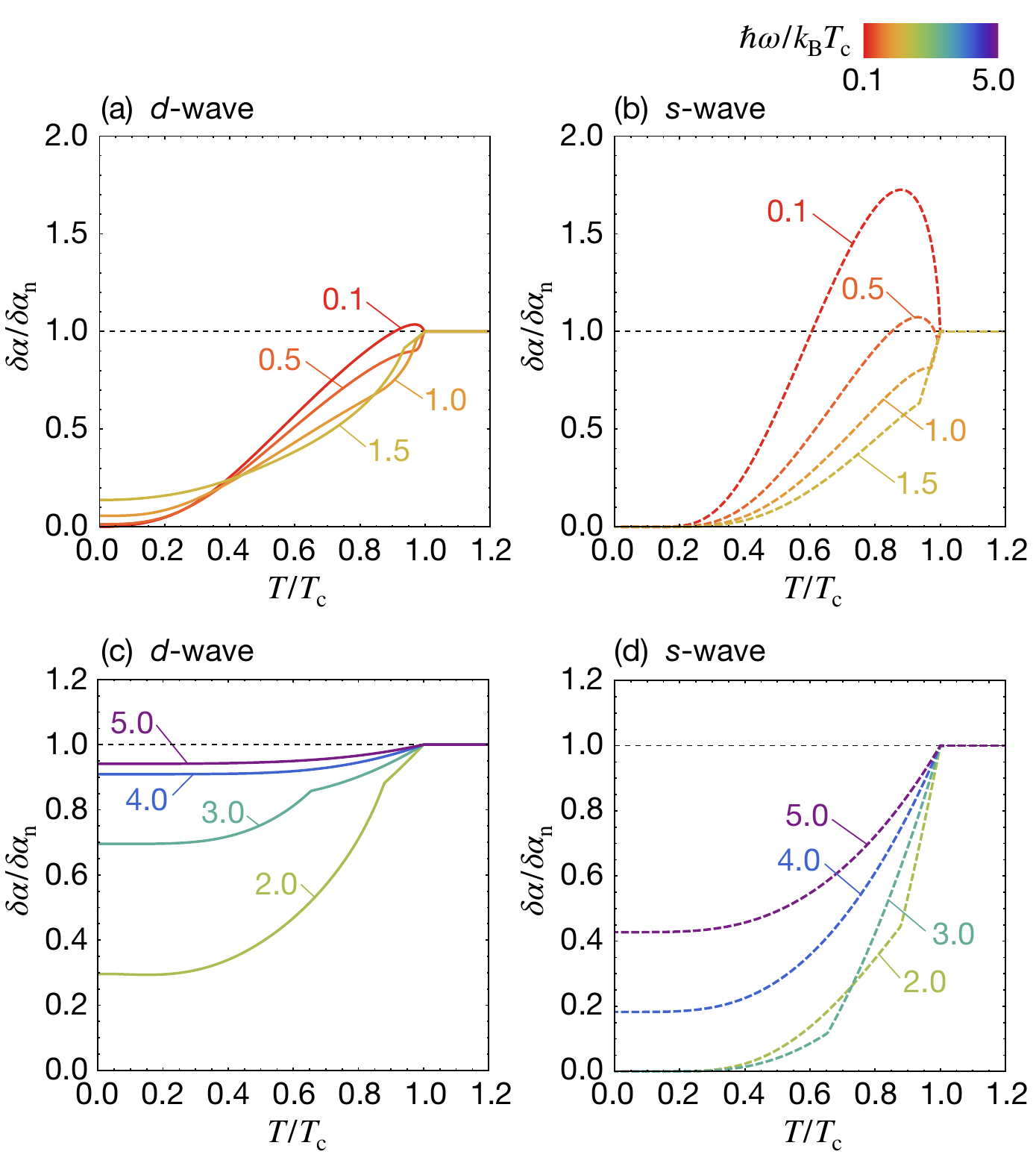}
\end{center}
\caption{
Enhanced Gilbert damping $\delta\alpha$ as a function of temperature $T$.
The left panels (a) and (c) show $\delta\alpha$ in the $d$-wave SC in the low and high frequency cases, respectively.
The right panels (b) and (d) show $\delta\alpha$ in the $s$-wave SC in the low and high frequency cases, respectively.
$\delta\alpha_{\mathrm{n}}$ is the normal state value.
}
\label{fig_T_dep}
\end{figure}

Figure \ref{fig_T_dep} shows the enhanced Gilbert damping constant $\delta\alpha$ as a function of temperature for several FMR frequencies, where
$\delta\alpha$ is normalized with respect to its value in the normal state.
We compare $\delta\alpha$ in the $d$-wave SC shown in Figs.~\ref{fig_T_dep}~(a) and (c) to that in the $s$-wave SC shown in Figs.~\ref{fig_T_dep}~(b) and (d). 
	The enhanced Gilbert damping for the $s$-wave SC is given by \cite{Kato2019}
\begin{align}
	\delta\alpha &= \frac{2\pi S {J_2}^2 D_{\mathrm{n}}^2}{\hbar\omega}
	\int_{-\infty}^{\infty} dE [f(E)-f(E+\hbar\omega)]
	\notag\\& \quad \times
	\left(
	 1 + \frac{\Delta^2}{E(E+\hbar\omega)}
	\right)
	\notag\\&\quad
	\times \mathrm{Re} \left[\frac{|E|}{\sqrt{E^2-\Delta^2}}\right]
	\mathrm{Re} \left[\frac{|E+\hbar\omega|}{\sqrt{(E+\hbar\omega)^2-\Delta^2}}\right],
\end{align}
where the temperature dependence of $\Delta$ is the same as that for the $d$-wave SC, given by Eq.~(\ref{Delta_T}).
Note that the BCS theory we are based on, which is valid when the Fermi energy is much larger than $\Delta$, is described by only some universal parameters, including $T_{\mathrm{c}}$, and independent of the detail of the system in the normal state.
When $\hbar\omega/k_{\mathrm{B}}T_{\mathrm{c}}=0.1$, $\delta\alpha$ shows a coherence peak just below the transition temperature $T_{\mathrm{c}}$.
However, the coherence peak of the $d$-wave SC is smaller than that of the $s$-wave SC.
Within the low temperature limit, $\delta\alpha$ in the $d$-wave SC shows power-law decay behavior described by $\delta\alpha\propto T^2$.
This is in contrast to $\delta\alpha$ in the $s$-wave SC, which shows exponential decay.
The difference in the low temperature region originates from the densities of states in the $d$-wave and $s$-wave SCs, which have gapless and full gap structures, respectively.
When the FMR frequency increases, the coherence peak is suppressed, and $\delta\alpha$ decays monotonically with decreasing temperature.
$\delta\alpha$ has a kink structure at $\hbar\omega=2\Delta$, where the FMR frequency corresponds to the superconducting gap.

\begin{figure}
\begin{center}
\includegraphics[width=1\hsize]{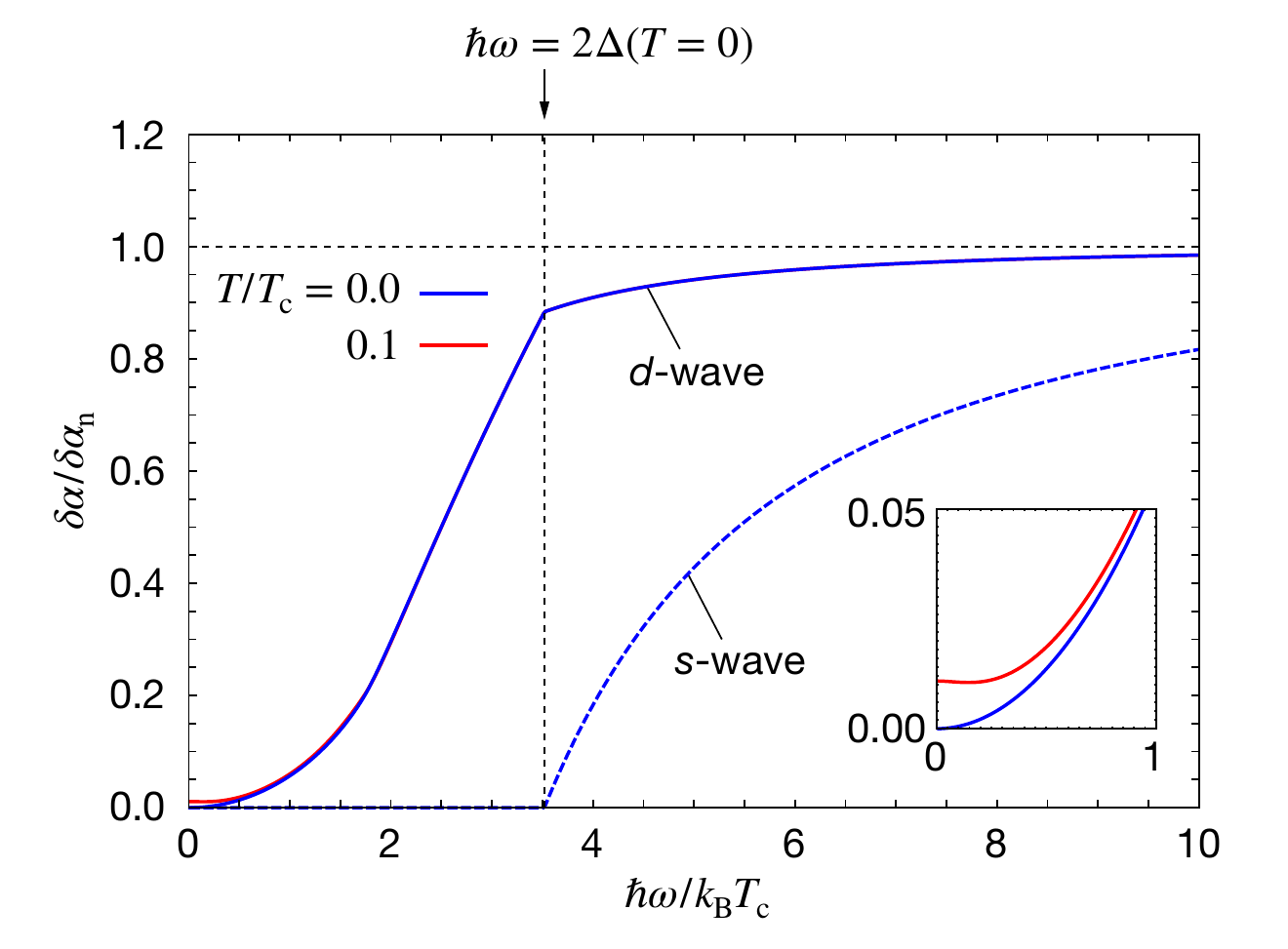}
\end{center}
\caption{
Enhanced Gilbert damping $\delta\alpha$ as a function of frequency $\omega$. The vertical dotted line indicates the resonance frequency $\hbar\omega=2\Delta(T=0)$. The inset shows an enlarged view in the low-frequency region.
}
\label{fig_omega_dep}
\end{figure}

Figure \ref{fig_omega_dep} shows $\delta\alpha$ at $T=0$ as a function of $\omega$.
In the $d$-wave SC, $\delta\alpha$ grows from zero with increasing $\omega$ as $\delta\alpha\propto\omega^2$.
When the value of $\delta\alpha$ becomes comparable to the normal state value, the increase in $\delta\alpha$ is suppressed, and $\delta\alpha$ then approaches the value in the normal state.
In contrast, $\delta\alpha$ in the $s$-wave SC vanishes as long as the condition that $\hbar\omega<2\Delta$ is satisfied.
When $\hbar\omega$ exceeds $2\Delta$, $\delta\alpha$ then increases with increasing $\omega$ and approaches the normal state value.
This difference also originates from the distinct spectral functions of the $d$-wave and $s$-wave SCs.
Under the low temperature condition that $T=0.1T_{\rm c}$, the frequency dependence of $\delta \alpha$ does not change for the $s$-wave SC, and it only changes in the low-frequency region where $\hbar \omega \lesssim k_{\rm B}T$ for the $d$-wave SC (see the inset in Fig.~\ref{fig_omega_dep}).

\section{Discussion}
\label{sec_discussion}

We discuss the modulation of the Gilbert damping in SCs with nodes other than the $d$-wave SC considered in this work. Other SCs with nodes are expected to exhibit the power-law decay behavior within the low-temperature and low-frequency limit as the $d$-wave SCs. However, the exponent of the power can differ due to the difference of the quasiparticle density of states. Furthermore, in the $p$-wave states, two significant differences arise due to spin-triplet Cooper pairs. First, the uniform spin susceptibility $\chi_{\vq=\bm{0}}^R(\omega)$ can be finite in the spin-triplet SCs because the spin is not conserved. Second, the enhanced Gilbert damping exhibits anisotropy and the value changes by changing the relative angle between the Cooper pair spin and localized spin \cite{ominato2021detection}.

In our work, proximity effect between FIs and SCs was not taken into account because the FMR modulation was calculated by second-order perturbation based on the tunnel Hamiltonian.
Reduction of superconducting gap due to the proximity effect~\cite{silaevLargeEnhancementSpin2020} and effect of the subgap Andreev bound states that appear in the $ab$-axis junction~\cite{TanakaKashiwaya1995} would also be an important problem left for future works.

Physics of the FMR modulation for $d$-wave SC/ferromagnetic metal junctions is rather different from that for $d$-wave SC/FI junctions. For $d$-wave SC/ferromagnetic metal junctions, spin transport is described by electron hopping across a junction and the FMR modulation is determined by the product of the density of states of electrons for a $d$-wave SC and a ferromagnetic metal. (We note that the FMR modulation is determined by a spin susceptibility of $d$-wave SC, which in general includes different information from the density of states of electrons.) While the FMR modulation is expected to be reduced below a SC transition temperature due to opening an energy gap, its temperature dependence would be different from results obtained in our work.

Finally, let us discuss effect of the diagonal exchange coupling given in Eq.~(\ref{eq:Hzzcoupling})
(see also the last part of Sec.~\ref{sec_model}).
This term causes an exchange bias, i.e., an effective Zeeman field on conduction electrons in the $d$-wave SC, which is derived as follows.
First, the $x$-component of the localized spin is approximated as $\langle S^x_j \rangle \approx S$, which gives $S^x_{\bm{k}}\approx S\sqrt{N}\delta_{\bm{k},\bm{0}}$.
Next, the matrix element $J_{\bm{q},\bm{k}=\bm{0}}$ is replaced by the configurationally averaged value $J_{\bm{q},\bm{k}=\bm{0}}=J_1\delta_{\bm{q},\bm{0}}$.
Consequently, the effective Zeeman field term is given by
\begin{align}
H_{\mathrm{Z}} \approx E_{\mathrm{Z}} \sum_{\bm p} (c^\dagger_{{\bm p}\uparrow} c_{{\bm p}\uparrow} - c^\dagger_{{\bm p}\downarrow} c_{{\bm p}\downarrow}),
\end{align}
where we introduced a Zeeman energy as $E_{\mathrm{Z}} = J_1S\sqrt{N}$.
This term induces spin splitting of conduction electrons in the $d$-wave SC and changes the spin susceptibility of the SC.
The spin-splitting effect causes a spin excitation gap and modifies the frequency dependence in Fig.~\ref{fig_omega_dep}, that will provide additional information on the exchange coupling at the interface.
In actual experimental setup for the $d$-wave SC, however, the Zeeman energy, that is less than the exchange bias between a magnetic insulator and a metal, is estimated to be of the order of $0.1 \, {\rm erg}/{\rm cm}^2$.
This leads to the exchange coupling that is much less than $J\sim 0.1 \, {\rm meV}$ for YIG~\cite{NOGUES1999203}.
Therefore, we expect that the interfacial exchange coupling is much smaller than the superconducting gap and the microwave photon energy though it has not been measured so far.
A detailed analysis for this spin-splitting effect is left for a future problem.

\section{Conclusion}
\label{sec_conclusion}

In this work, we have investigated Gilbert damping modulation in the $d$-wave SC/FI bilayer system. The enhanced Gilbert damping constant in this case is proportional to the imaginary part of the dynamic spin susceptibility of the $d$-wave SC.
We found that the Gilbert damping modulation reflects the gapless excitation that is inherent in $d$-wave SCs.
The coherence peak is suppressed in the $d$-wave SC when compared with that in the $s$-wave SC.
In addition, the differences in the spectral functions for the $d$-wave and $s$-wave SCs with gapless and full-gap structures lead to power-law and exponential decays within the low-temperature limit, respectively.
Within the low-temperature limit, $\delta\alpha$ in the $d$-wave SC increases with increasing $\omega$, while $\delta\alpha$ in the $s$-wave SC remains almost zero as long as the excitation energy $\hbar\omega$ remains smaller than the superconducting gap $2\Delta$.

Our results illustrate the usefulness of measurement of the FMR modulation of unconventional SCs for determination of their symmetry through spin excitation.
We hope that this fascinating feature will be verified experimentally in $d$-wave SC/FI junctions in the near future.
To date, one interesting result of FMR modulation in $d$-wave SC/ferromagnetic metal structures has been reported~\cite{carreira2020spin}.
This modulation can be dependent on metallic states, which are outside the scope of the theory presented here.
The FMR modulation caused by ferromagnetic metals is another subject that will have to be clarified theoretically in future work.

Furthermore, our work provides the most fundamental basis for application to analysis of junctions with various anisotropic SCs.
For example, some anisotropic SCs are topological and have an intrinsic gapless surface state. SP can be accessible and can control the spin excitation of the surface states because of its interface sensitivity. The extension of SP to anisotropic and topological superconductivity represents one of the most attractive directions for further development of superconducting spintronics.

{\it Acknowledgments.---}
This work is partially supported by the Priority Program of Chinese Academy of Sciences, Grant No. XDB28000000.
We acknowledge JSPS KAKENHI for Grants (No. JP20H01863, No. JP20K03835, No. JP20K03831, No. JP20H04635, and No.21H04565).

\appendix

\section{Magnon self-energy induced by a rough interface}
\label{roughness}

The roughness of the interface is taken into account as an uncorrelated (white noise) distribution of the exchange couplings \cite{ominato2021detection}, as shown below.  
We start with an exchange model in the real space
\begin{align}
 H_{\mathrm{ex}} &= \sum_j \int d^2r J(\boldsymbol{r},  \boldsymbol{r}_j) \boldsymbol{\sigma}(\boldsymbol{r}) \cdot \boldsymbol{S}_j
 \notag\\
 &= \sum_{\boldsymbol{q}, \boldsymbol{k}} J_{\boldsymbol{q}, \boldsymbol{k}} \boldsymbol\sigma_{\boldsymbol{q}} \cdot \boldsymbol{S}_{\boldsymbol{k}}.
\end{align}
The spin density $\boldsymbol{\sigma}(\vr)$ in the SC and the spin $\boldsymbol{S}_j$ in the FI are represented in the momentum space as 
\begin{align}
 \boldsymbol{\sigma}(\boldsymbol{r}) 
 &= \frac{1}{A} \sum_{\boldsymbol{q}} e^{i \boldsymbol{q} \cdot \boldsymbol{r}} \boldsymbol{\sigma}_{\boldsymbol{q}},
 \\
 \boldsymbol{S}_j
 &=
 \frac{1}{\sqrt{N}}
 \sum_{\boldsymbol{k}} 
 e^{i \boldsymbol{k} \cdot \boldsymbol{r}_j} \boldsymbol{S}_{\boldsymbol{k}},
\end{align}
where $A$ denotes the area of the system and $N$ is the number of sites.
The exchange coupling constant is also obtained to be
\begin{align}
 J_{\boldsymbol{q}, \boldsymbol{k}}
 = \frac{1}{A\sqrt{N}}
 \sum_j\int d^2r e^{i(\boldsymbol{q} \cdot \boldsymbol{r} + \boldsymbol{k} \cdot \boldsymbol{r}_j)}
 J(\boldsymbol{r}, \boldsymbol{r}_j).
\end{align}
The exchange model $H_{\mathrm{ex}}$ is decomposed into the spin transfer term $H_{\mathrm{T}}$ and the effective Zeeman field term $H_{\mathrm{Z}}$ as $H_{\mathrm{ex}}=H_{\mathrm{T}}+H_{\mathrm{Z}}$.

Now we consider the roughness effect of the interface.
Uncorrelated roughness is expressed by the mean $J_1$ and variance ${J_2}^2$ as
\begin{align}
 \frac{1}{\sqrt{N}}\sum_j
 \overline{J(\boldsymbol{r}, \boldsymbol{r}_j)} 
 &= J_1,
 \\
 \frac{1}{N}\sum_{jj'}
 \overline{J(\boldsymbol{r}, \boldsymbol{r}_j)
 J(\boldsymbol{r}', \boldsymbol{r}_{j'})} - {J_1}^2
 &= {J_2}^2 A \delta^2(\boldsymbol{r}-\boldsymbol{r}'),
\end{align}
where $\overline{O}$ is the configurational average of $O$ over the roughness. 
The above expressions lead to the configurationally averaged self-energy
\begin{align}
\Sigma^R_{\boldsymbol{k}=\boldsymbol{0}}(\omega)
 &=- \sum_{\boldsymbol{q}}
 \overline{|J_{\boldsymbol{q}, \vk=\boldsymbol{0}}|^2}
 \chi_{\vq}^R(\omega)
 \notag\\&
 = -{J_1}^2 \chi_{\boldsymbol{q} = \boldsymbol{0}}^R(\omega)
 - {J_2}^2 \sum_{\boldsymbol{q}} \chi_{\vq}^R(\omega),
\end{align}
which coincides with the model Eq.~(\ref{Jqk}) in the main text.
This model provides a smooth connection between the specular (${J_1}^2 \chi_{\boldsymbol{q}=\boldsymbol{0}}^R$) and diffuse (${J_2}^2 \sum_{\boldsymbol{q}} \chi_{\boldsymbol{q}}^R$) limits. 
The uncorrelated roughness case introduced above is a simple linear interpolation of the two. Extensions to correlated roughness can be made straightforwardly.

\bibliography{ref}

\end{document}